\documentstyle[12pt,epsf]{article}
\newcommand{\be}{\begin{equation}}
\newcommand{\ee}{\end{equation}}

\newcommand{\beqa}{\begin{eqnarray}}
\newcommand{\eeqa}{\end{eqnarray}}
\newcommand{\nn}{\nonumber}

\newcommand{\eqref}[1]{(\ref{#1})}

\def\boxit#1{\vbox{\hrule\hbox{\vrule\kern8pt
\vbox{\hbox{\kern8pt}\hbox{\vbox{#1}}\hbox{\kern8pt}}
\kern8pt\vrule}\hrule}}
\def\mathboxit#1{\vbox{\hrule\hbox{\vrule\kern8pt\vbox{\kern8pt
\hbox{$\displaystyle #1$}\kern8pt}\kern8pt\vrule}\hrule}}

\def\IB{\relax\hbox{$\inbar\kern-.3em{\rm B}$}}
\def\IC{\relax\hbox{$\inbar\kern-.3em{\rm C}$}}
\def\ID{\relax\hbox{$\inbar\kern-.3em{\rm D}$}}
\def\IE{\relax\hbox{$\inbar\kern-.3em{\rm E}$}}
\def\IF{\relax\hbox{$\inbar\kern-.3em{\rm F}$}}
\def\IG{\relax\hbox{$\inbar\kern-.3em{\rm G}$}}
\def\IGa{\relax\hbox{${\rm I}\kern-.18em\Gamma$}}
\def\IH{\relax{\rm I\kern-.18em H}}
\def\IK{\relax{\rm I\kern-.18em K}}
\def\IL{\relax{\rm I\kern-.18em L}}
\def\IP{\relax{\rm I\kern-.18em P}}
\def\IR{\relax{\rm I\kern-.18em R}}
\def\IZ{\relax\ifmmode\mathchoice
{\hbox{\cmss Z\kern-.4em Z}}{\hbox{\cmss Z\kern-.4em Z}}
{\lower.9pt\hbox{\cmsss Z\kern-.4em Z}} {\lower1.2pt\hbox{\cmsss
Z\kern-.4em Z}}\else{\cmss Z\kern-.4em Z}\fi}

\def\II{\relax{\rm I\kern-.18em I}}

\def\CA {{\cal A}}

\def\CG {{\cal G}}
\def\CH {{\cal H}}

\def\CL {{\cal L}}

\def\CP {{\cal P}}

\def\CU {{\cal U}}

\begin{document}

\hfill  NRCPS-HE-10-10

\vspace{1cm}
\begin{center}
{\Large ~\\{\it Non-Abelian Tensor Gauge Fields
}

}

\vspace{2cm}

{\sl George Savvidy\\
Demokritos National Research Center\\
Institute of Nuclear Physics\\
Ag. Paraskevi, GR-15310 Athens,Greece  \\
}
\end{center}
\vspace{2cm}

\centerline{{\bf Abstract}}

\vspace{12pt}

\noindent

Recently proposed extension of Yang-Mills theory contains non-Abelian
tensor gauge fields. The Lagrangian has quadratic kinetic terms, as well as cubic and
quartic terms describing non-linear interaction of tensor gauge fields
with the dimensionless coupling constant. We analyze particle content of
non-Abelian tensor gauge fields.  In four-dimensional space–time the
rank-2 gauge field describes propagating modes of helicity 2 and 0.
We introduce interaction of the non-Abelian tensor gauge field
with fermions and demonstrate that the free equation of motion for the spin-vector field
correctly describes the propagation of massless modes of helicity 3/2.
We have found a new metric-independent gauge invariant
density  which is a four-dimensional analog of the
Chern-Simons density. The Lagrangian augmented by this Chern-Simons-like invariant
describes massive Yang-Mills boson,
providing a gauge-invariant mass gap for a four-dimensional gauge field theory.

 \vspace{20 pt}

\centerline{\it Invited talk given at "Gauge Fields. Yesterday, Today, Tomorrow"}
\centerline{\it Moscow 2010,  in Honor of the 70-th Birthday of Andrey Slavnov}


\newpage

\pagestyle{plain}


\section{\it Introduction}

It is appealing to extend the Yang-Mills theory \cite{yang,chern} so that it will define the
interaction of  fields which carry not only non-commutative internal charges, but
also arbitrary  large spins. This extension
will induce the interaction of matter fields mediated by charged
gauge quanta carrying spin larger than one \cite{Savvidy:2005fi}.
In our recent approach these gauge fields are defined as
rank-$(s+1)$ tensors
\cite{Savvidy:2005fi,Savvidy:2005zm,Savvidy:2005ki,Barrett:2007nn}
$$
A^{a}_{\mu\lambda_1 ... \lambda_{s}}(x)
$$
and are totally symmetric with respect to the
indices $  \lambda_1 ... \lambda_{s}  $.
The index $s$ runs from zero to infinity.
The first member of this family of the tensor gauge bosons is the Yang-Mills
vector boson $A^{a}_{\mu}$.

The extended non-Abelian gauge transformation $\delta_{\xi}$ of the tensor gauge fields
comprises a closed algebraic structure \cite{Savvidy:2005fi,Savvidy:2005zm,Savvidy:2005ki}.
This allows to define generalized field strength tensors
$
G^{a}_{\mu\nu,\lambda_{1}...\lambda_{s}}
$,
which are {\it transforming homogeneously}
with respect to the extended gauge transformations $\delta_{\xi}$.
The field strength tensors
are used to construct two infinite series of gauge invariant quadratic forms
$
{{\cal L}}_{s}$ and ${{\cal L}}^{'}_{s}
$.
These forms contain quadratic kinetic terms  and terms  describing
nonlinear interaction of Yang-Mills type.
In order to make all tensor gauge fields dynamical one should add all these
forms in the Lagrangian
\cite{Savvidy:2005fi,Savvidy:2005zm,Savvidy:2005ki}.

The fermions are defined as Rarita-Schwinger spinor-tensors \cite{rarita,singh1,fronsdal1}
$$
\psi^{\alpha}_{\lambda_1 ... \lambda_{s}}(x)
$$
with mixed transformation properties of Dirac four-component wave
function (the index $\alpha$ denotes the Dirac index) and
are totally symmetric tensors of the rank $s$
over the indices $\lambda_1 ... \lambda_{s}$.
All fields of the $\{ \psi \}$ family
are isotopic multiplets belonging to the
same representation $\sigma$ of the compact Lie group G
(the corresponding indices are suppressed).
The gauge invariant Lagrangian for fermions
also contains a linear sum of two infinite series of forms
${{\cal L}}_{s +1/2}$ and ${{\cal L}}^{'}_{s+1/2 }$.
The coupling constants in front of these forms
remain arbitrary because all terms in the sum are separately gauge invariant.
The extended gauge symmetry alone does not define them.
{\it The basic principle which we shall pursue
in our construction will be to fix these coupling constants demanding
unitarity of the theory}\footnote{
For that one should study the spectrum of the theory and to prove that there are
no propagating negative norm states, that is, ghost states.}.

In the second and third sections we shall outline the transformation properties
of non-Abelian tensor gauge fields, the definition of
the corresponding field stress tensors, the general expression for the invariant Lagrangian
and the description of propagating modes for the lower rank
tensor gauge fields \cite{Savvidy:2005fi,Savvidy:2005zm,Savvidy:2005ki,Savvidy:2009zz}.
In the forth and fifths sections we shall incorporate into the theory fermions of
half-integer spins \cite{Savvidy:2008zy}.  We shall construct two infinite series of gauge
invariant forms and study the propagating modes for lower rank fermion fields.

In the sixth sections we shall construct a metric-independent gauge invariant
density  which is a four-dimensional analog of the
Chern-Simons density. The Lagrangian augmented by this Chern-Simons-like invariant
describes massive Yang-Mills boson, providing a gauge-invariant mass gap for a
four-dimensional gauge field theory \cite{Savvidy:2010bk}.

\section{\it Non-Abelian Tensor Gauge Fields}

The gauge fields are defined as rank-$(s+1)$ tensors
\cite{Savvidy:2005fi}
$$
A^{a}_{\mu\lambda_1 ... \lambda_{s}}(x),~~~~~s=0,1,2,...
$$
and are totally symmetric with respect to the
indices $  \lambda_1 ... \lambda_{s}  $.  A priory the tensor fields
have no symmetries with
respect to the first index  $\mu$. The index $a$ numerates the generators $L^a$
of the Lie algebra $\breve{g}$ of a {\it compact}\footnote{The algebra $\breve{g}$
possesses an orthogonal
basis in which the structure constants $f^{abc}$ are totally antisymmetric.}
Lie group G. One can think of these tensor fields as appearing in the
expansion of the extended gauge field $\CA_{\mu}(x,e)$ over the unite  vector
$e_{\lambda}$
\cite{Savvidy:2005ki}:
\be\label{gaugefield}
{\cal A}_{\mu}(x,e)=\sum_{s=0}^{\infty} {1\over s!} ~A^{a}_{\mu\lambda_{1}...
\lambda_{s}}(x)~L^{a}e_{\lambda_{1}}...e_{\lambda_{s}}.
\ee
The gauge field $A^{a}_{\mu\lambda_1 ... \lambda_{s}}$ carries
indices $a,\lambda_1, ..., \lambda_{s}$ labeling the generators of {\it extended current
algebra $\CG$ associated with compact Lie group G.} It has infinitely many generators
$L^{a}_{\lambda_1 ... \lambda_{s}} = L^a e_{\lambda_1}...e_{\lambda_s}$ and
the corresponding algebra is given by the commutator \cite{Savvidy:2005ki}
\be
[L^{a}_{\lambda_1 ... \lambda_{i}}, L^{b}_{\lambda_{i+1} ... \lambda_{s}}]=if^{abc}
L^{c}_{\lambda_1 ... \lambda_{s} }.
\ee
Because $L^{a}_{\lambda_1 ... \lambda_{s}}$ are
space-time tensors, the full algebra includes the Poincar\'e generators $P_{\mu},~M_{\mu\nu}$.
They act on the space-time components of the above generators as follows \cite{Savvidy:2008zy}:
\beqa\label{extensionofpoincarealgebra}
~&&[P^{\mu},~P^{\nu}]=0,\nn\\
~&&[M^{\mu\nu},~P^{\lambda}] = i(\eta^{\lambda \nu}~P^{\mu}
- \eta^{\lambda \mu }~P^{\nu}) ,\nn\\
~&&[M^{\mu \nu}, ~ M^{\lambda \rho}] = i(\eta^{\mu \rho}~M^{\nu \lambda}
-\eta^{\mu \lambda}~M^{\nu \rho} +
\eta^{\nu \lambda}~M^{\mu \rho}  -
\eta^{\nu \rho}~M^{\mu \lambda} ),\nonumber\\
~&&[P^{\mu},~L_{a}^{\lambda_1 ... \lambda_{s}}]=0, \nn\\
~&&[M^{\mu \nu}, ~ L_{a}^{\lambda_1 ... \lambda_{s}}] = i(
\eta^{\lambda_1\nu } L_{a}^{\mu \lambda_2... \lambda_{s}}
-\eta^{\lambda_1\mu} L_{a}^{\nu\lambda_2... \lambda_{s}}
+...+
\eta^{\lambda_s\nu } L_{a}^{\lambda_1... \lambda_{s-1}\mu } -
\eta^{\lambda_s\mu } L_{a}^{\lambda_1... \lambda_{s-1}\nu } ),\nonumber\\
~&&[L_{a}^{\lambda_1 ... \lambda_{i}}, L_{b}^{\lambda_{i+1} ... \lambda_{s}}]=if_{abc}~
L_{c}^{\lambda_1 ... \lambda_{s} } ,     ~~~(\mu,\nu,\rho,\lambda=0,1,2,3; ~~~~~s=0,1,2,... )
\eeqa
It is an infinite-dimensional extension of the Poincar\'e algebra by generators which contains
isospin algebra G. In some sense the new vector variable $e_\lambda$ plays  a role
similar to the Grassmann variable $\theta$ in supersymmetry algebras \cite{Coleman:1967ad,Haag:1974qh}.

{\it The extended non-Abelian gauge transformations of the
tensor gauge fields are defined
by the following equations } \cite{Savvidy:2005zm}:
\beqa\label{polygauge}
\delta A^{a}_{\mu} &=& ( \delta^{ab}\partial_{\mu}
+g f^{acb}A^{c}_{\mu})\xi^b ,~~~~~\\
\delta A^{a}_{\mu\nu} &=&  ( \delta^{ab}\partial_{\mu}
+  g f^{acb}A^{c}_{\mu})\xi^{b}_{\nu} + g f^{acb}A^{c}_{\mu\nu}\xi^{b},\nonumber\\
\delta A^{a}_{\mu\nu \lambda}& =&  ( \delta^{ab}\partial_{\mu}
+g f^{acb} A^{c}_{\mu})\xi^{b}_{\nu\lambda} +
g f^{acb}(  A^{c}_{\mu  \nu}\xi^{b}_{\lambda } +
A^{c}_{\mu \lambda }\xi^{b}_{ \nu}+
A^{c}_{\mu\nu\lambda}\xi^{b}),\nn\\
.........&.&............................ ,\nn
\eeqa
where $\xi^{a}_{\lambda_1 ... \lambda_{s}}(x)$ are totally symmetric gauge parameters.
These extended gauge transformations
generate a closed algebraic structure. In order to see that one should compute the
commutator of two extended gauge transformations $\delta_{\eta}$ and $\delta_{\xi}$
of parameters $\eta$ and $\xi$.
The commutator of two transformations can be expressed in the following form \cite{Savvidy:2005zm}:
\be\label{gaugecommutator}
[~\delta_{\eta},\delta_{\xi}]~A_{\mu\lambda_1\lambda_2 ...\lambda_s} ~=~
-i g~ \delta_{\zeta} A_{\mu\lambda_1\lambda_2 ...\lambda_s}
\ee
and is again an extended gauge transformation with the gauge parameters
$\{\zeta\}$ which are given by the matrix commutators
\beqa\label{gaugealgebra}
\zeta&=&[\eta,\xi]\\
\zeta_{\lambda_1}&=&[\eta,\xi_{\lambda_1}] +[\eta_{\lambda_1},\xi]\nn\\
\zeta_{\nu\lambda} &=& [\eta,\xi_{\nu\lambda}] +  [\eta_{\nu},\xi_{\lambda}]
+ [\eta_{\lambda},\xi_{\nu}]+[\eta_{\nu\lambda},\xi],\nn\\
......&.&..........................\nn
\eeqa
Each single field $A^{a}_{\mu \lambda_1...\lambda_s}(x),~s=1,2,3,...$ has no geometrical interpretation,
but their union has a geometrical interpretation in terms of {\it connection on
the extended vector bundle X} \cite{Savvidy:2005ki}. Indeed,
one can define the extended vector bundle X whose structure group
is $\CG$ with group elements
$
U(\xi)=exp[~i \xi(x,e)~],
$
where
$$
\xi(x,e)=  \sum_s {1\over s!}~\xi^{a}_{\lambda_1 ... \lambda_{s}}(x) ~~L^{a}e_{\lambda_{1}}...e_{\lambda_{s}}.
$$
Defining the extended gauge transformation of $\CA_{\mu}(x,e)$  in a standard way
\be\label{extendedgaugetransformation}
\CA^{'}_{\mu}(x,e) = U(\xi)  \CA_{\mu}(x,e) U^{-1}(\xi) -{i\over g}
\partial_{\mu}U(\xi) ~U^{-1}(\xi),
\ee
we get the extended vector bundle X on which the gauge field
$\CA^{a}_{\mu}(x,e)$ is a connection \cite{chern}.
The expansion of (\ref{extendedgaugetransformation}) over the vector
$e_\lambda$ reproduces gauge transformation law of the tensor gauge  fields
(\ref{polygauge}). Using the commutator of the covariant derivatives
$
\nabla^{ab}_{\mu} = (\partial_{\mu}-ig \CA_{\mu}(x,e))^{ab}
$
\be
[\nabla_{\mu}, \nabla_{\nu}]^{ab} = g f^{acb} \CG^{c}_{\mu\nu}~,
\ee
we can define the extended field strength tensor
\be\label{fieldstrengthgeneral}
\CG_{\mu\nu}(x,e) = \partial_{\mu} \CA_{\nu}(x,e) - \partial_{\nu} \CA_{\mu}(x,e) -
i g [ \CA_{\mu}(x,e)~\CA_{\nu}(x,e)]
\ee
which transforms homogeneously:
$
\CG^{'}_{\mu\nu}(x,e)) = U(\xi)  \CG_{\mu\nu}(x,e) U^{-1}(\xi).
$
Thus the {\it generalized field strengths  are defined as} \cite{Savvidy:2005zm}
\beqa\label{fieldstrengthparticular}
G^{a}_{\mu\nu} &=&
\partial_{\mu} A^{a}_{\nu} - \partial_{\nu} A^{a}_{\mu} +
g f^{abc}~A^{b}_{\mu}~A^{c}_{\nu},\\
G^{a}_{\mu\nu,\lambda} &=&
\partial_{\mu} A^{a}_{\nu\lambda} - \partial_{\nu} A^{a}_{\mu\lambda} +
g f^{abc}(~A^{b}_{\mu}~A^{c}_{\nu\lambda} + A^{b}_{\mu\lambda}~A^{c}_{\nu} ~),\nn\\
G^{a}_{\mu\nu,\lambda\rho} &=&
\partial_{\mu} A^{a}_{\nu\lambda\rho} - \partial_{\nu} A^{a}_{\mu\lambda\rho} +
g f^{abc}(~A^{b}_{\mu}~A^{c}_{\nu\lambda\rho} +
 A^{b}_{\mu\lambda}~A^{c}_{\nu\rho}+A^{b}_{\mu\rho}~A^{c}_{\nu\lambda}
 + A^{b}_{\mu\lambda\rho}~A^{c}_{\nu} ~),\nn\\
 ......&.&............................................\nn
\eeqa
and transform homogeneously with respect to the extended
gauge transformations (\ref{polygauge}). The field strength tensors are
antisymmetric in their first two indices and are totally symmetric with respect to the
rest of the indices.
The inhomogeneous extended gauge transformation (\ref{polygauge})
induces the homogeneous gauge
transformation of the corresponding field strength tensors
(\ref{fieldstrengthparticular}) of the form \cite{Savvidy:2005zm}
\beqa\label{fieldstrenghparticulartransformation}
\delta G^{a}_{\mu\nu}&=& g f^{abc} G^{b}_{\mu\nu} \xi^c  ,\\
\delta G^{a}_{\mu\nu,\lambda} &=& g f^{abc} (~G^{b}_{\mu\nu,\lambda} \xi^c
+ G^{b}_{\mu\nu} \xi^{c}_{\lambda}~),\nonumber\\
\delta G^{a}_{\mu\nu,\lambda\rho} &=& g f^{abc}
(~G^{b}_{\mu\nu,\lambda\rho} \xi^c
+ G^{b}_{\mu\nu,\lambda} \xi^{c}_{\rho} +
G^{b}_{\mu\nu,\rho} \xi^{c}_{\lambda} +
G^{b}_{\mu\nu} \xi^{c}_{\lambda\rho}~),\nn\\
......&.&..........................\nn
\eeqa
Using these field strength tensors one can
construct two infinite series of forms
$
{{\cal L}}_{s}$ and ${{\cal L}}^{'}_{s}~(s=2,3,...)
$
invariant with respect to the
transformations $\delta_{\xi} $. They are quadratic in field strength
tensors. The
first series is given by the formula \cite{Savvidy:2005fi,Savvidy:2005zm,Savvidy:2005ki}
\beqa\label{fulllagrangian1}
{{\cal L}}_{s+1}&=&-{1\over 4} ~
G^{a}_{\mu\nu, \lambda_1 ... \lambda_s}~
G^{a}_{\mu\nu, \lambda_{1}...\lambda_{s}} +.......\nonumber\\
&=& -{1\over 4}\sum^{2s}_{i=0}~a^{s}_i ~
G^{a}_{\mu\nu, \lambda_1 ... \lambda_i}~
G^{a}_{\mu\nu, \lambda_{i+1}...\lambda_{2s}}
(\sum_{P } \eta^{\lambda_{i_1} \lambda_{i_2}} .......
\eta^{\lambda_{i_{2s-1}} \lambda_{i_{2s}}})~,
\eeqa
where the sum $\sum_P$ runs over all nonequal permutations of
$\lambda_i~'s$, in total $(2s-1)!!$
terms, and the numerical coefficients are $a^{s}_i = {s!\over i!(2s-i)!}$.
The second series of gauge invariant quadratic forms is given by the formula
\cite{Savvidy:2005fi,Savvidy:2005zm,Savvidy:2005ki}
\beqa\label{secondfulllagrangian}
{{\cal L}}^{'}_{s+1}&=&{1\over 4} ~
G^{a}_{\mu\nu,\rho\lambda_3  ... \lambda_{s+1}}~
G^{a}_{\mu\rho,\nu\lambda_{3} ...\lambda_{s+1}} +{1\over 4} ~
G^{a}_{\mu\nu,\nu\lambda_3  ... \lambda_{s+1}}~
G^{a}_{\mu\rho,\rho\lambda_{3} ...\lambda_{s+1}} +.......\nonumber\\
&=& {1\over 4}\sum^{2s+1}_{i=1}~{ a^{s}_{i-1}\over s}  ~
G^{a}_{\mu\lambda_1,\lambda_2  ... \lambda_i}~
G^{a}_{\mu\lambda_{i+1},\lambda_{i+2} ...\lambda_{2s+2}}
(\sum^{'}_{P} \eta^{\lambda_{i_1} \lambda_{i_2}} .......
\eta^{\lambda_{i_{2s+1}} \lambda_{i_{2s+2}}})~,
\eeqa
where the sum $\sum^{'}_P$ runs over all nonequal permutations of
$\lambda_i~'s$, with exclusion
of the terms which contain $\eta^{\lambda_{1},\lambda_{i+1}}$.

These forms
contain quadratic {\it kinetic terms}, as well as cubic and
quartic terms  describing
{\it nonlinear interaction of gauge fields} with dimensionless
coupling constant $g$.
In order to make all tensor gauge fields dynamical one should add all these
forms in the Lagrangian
\cite{Savvidy:2005fi,Savvidy:2005zm,Savvidy:2005ki}:
\be\label{fulllagrangian3}
{{\cal L}} = {{\cal L}}_{YM} +   ({{\cal L}}_{2}+ {{\cal L}}^{'}_{2})
+g_{3}({{\cal L}}_{3}+ {4\over 3}{{\cal L}}^{'}_{3})+...+
g_{s+1}({{\cal L}}_{s+1}+ {2s\over s+1}{{\cal L}}^{'}_{s+1})+....
\ee
The coupling constants  $g_{3}, g_4,... $ remain arbitrary
because each term is separately invariant
with respect to the extended gauge transformations $\delta_\xi$
and leaves these coupling constants yet undetermined.

In the next section we shall analyze the free field equations for the lower rank
non-Abelian tensor gauge fields \cite{Savvidy:2009zz}. These equations are written in terms of the first
order derivatives of extended field strength tensors, similarly to the electrodynamics and the
Yang-Mills theory. In the Yang-Mills theory the free equation of motion
describes the propagation of massless gauge bosons of helicity $\lambda = \pm 1$.
The rank-2 gauge field describes propagating modes of helicity two and zero: $\lambda=\pm2,0$.
Thus the lower rank gauge fields have the following helicity content of propagating modes:
\beqa\label{helicities}
A_{\mu}:~~~~\lambda=\pm 1 ~,~~~~~~~~~~~~~~
A_{\mu\nu}:~~~~\lambda=\pm 2,0.
\eeqa
The propagating modes of higher rank gauge fields have been analyzed in \cite{Savvidy:2009zz}.

\section{\it Propagating Modes of Tensor Gauge Bosons}
In the Yang-Mills theory the free field equation is defined by the quadratic form:
\beqa\label{YMfreeoperator}
\CL_{YM}\vert_{quadratic} =
 {1 \over 2} A^{a}_{\alpha }
\CH_{\alpha \gamma }  A^{a}_{\gamma }, \nn
\eeqa
where
\be\label{YMfreeoperatorH}
\CH_{\alpha \gamma }=-k^2 \eta_{\alpha\gamma} + k_{\alpha}k_{\gamma}.
\ee
and describes the propagation of the massless gauge bosons of helicity $\lambda = \pm 1$:
$$
e_{\mu}^{\pm }= (0,1,\pm i,0).
$$
The kinetic term of the rank-2 gauge field is given by the quadratic form:
\be
\label{kineticterm}
{{\cal L}}_2 +   {{\cal L}}^{'}_2 ~\vert_{quadratic}  =  {1 \over 2} A^{a}_{\alpha\acute{\alpha}}
\CH_{\alpha\acute{\alpha}\gamma\acute{\gamma}}  A^{a}_{\gamma\acute{\gamma}},
\ee
where the kinetic operator is \cite{Savvidy:2005fi,Savvidy:2005zm,Savvidy:2005ki}
\beqa\label{quadraticform}
\CH_{\alpha\acute{\alpha}\gamma\acute{\gamma}}(k)=
(-\eta_{\alpha\gamma}\eta_{\acute{\alpha}\acute{\gamma}}
+{1 \over 2}\eta_{\alpha\acute{\gamma}}\eta_{\acute{\alpha}\gamma}
+{1 \over 2}\eta_{\alpha\acute{\alpha}}\eta_{\gamma\acute{\gamma}})k^2
+\eta_{\alpha\gamma}k_{\acute\alpha}k_{\acute{\gamma}}
+\eta_{\acute\alpha \acute{\gamma}}k_{\alpha}k_{\gamma}\nn\\
-{1 \over 2}(\eta_{\alpha\acute{\gamma}}k_{\acute\alpha}k_{\gamma}
+\eta_{\acute\alpha\gamma}k_{\alpha}k_{\acute{\gamma}}
+\eta_{\alpha\acute\alpha}k_{\gamma}k_{\acute{\gamma}}
+\eta_{\gamma\acute{\gamma}}k_{\alpha}k_{\acute\alpha}).
\eeqa
Thus one should solve the free equation of motion:
\be\label{basicequation}
\CH_{\alpha\acute{\alpha} \gamma\acute{\gamma}}(k)~
e_{\gamma\acute{\gamma}}(k) =0.
\ee
The vector space of independent solutions
$A_{\gamma\acute{\gamma}}=e_{\gamma\acute{\gamma}}(k) e^{ikx}$ depends on the rank  of the matrix
$\CH_{\alpha\acute{\alpha} \gamma\acute{\gamma}}(k)$.
Because the matrix operator $\CH_{\alpha\acute{\alpha} \gamma\acute{\gamma}}(k)$
explicitly depends on the momentum $k_{\mu}$, its $rank \CH=r(k)$ also depends on
momenta and therefore the number of independent solutions
${\cal N}$  depends on momenta
$
{\cal N}(k) = d-r(k)~.
$
The $rank~\CH$ is a Lorentz invariant quantity
and therefore depends on the value of momentum
square $k_{\mu}^2$.

The matrix operator (\ref{quadraticform})
in the four-dimensional space-time is a
$16 \times 16$ matrix\footnote{The multi-index $N \equiv
(\mu,\lambda)$ takes sixteen values.}.
In the reference frame, where
$k^{\gamma}=(\omega,0,0,k)$, it has a particularly simple form.
If $\omega^2 - k^2 \neq 0$, the rank of the 16-dimensional
matrix
$
H_{\alpha\acute{\alpha}\gamma\acute{\gamma}}(k)
$
is equal to $rank~H\vert_{\omega^2 - k^2 \ne 0}=9$
and the number of linearly independent solutions is $16-9=7$.
These seven solutions are  pure gauge fields
\be\label{puregaugepotentials}
e_{\gamma\acute{\gamma}}=
k_{\gamma}\xi_{\acute{\gamma}}+k_{\acute{\gamma}}\zeta_{\gamma},
\ee
where $\xi_{ \gamma }$ and $\zeta_{\gamma}$ are independent gauge parameters.
When $\omega^2 - k^2 = 0$, then the rank of the matrix
$
H_{\alpha\acute{\alpha}\gamma\acute{\gamma}}(k)
$
drops and   $rank~H  \vert_{\omega^2 - k^2 = 0}  =6$.
This leaves us with $16-6=10$ solutions. These are 7 solutions,
the pure gauge potentials (\ref{puregaugepotentials}),
and three new solutions representing the propagating modes:
\beqa\label{physicalmodes}
e_{\gamma\acute{\gamma}}^{\pm}= e_{\gamma}^{\pm}e_{ \acute{\gamma}}^{\pm}
 ,~~~~~
e_{\gamma\acute{\gamma}}^{A}=e_{\gamma}^{+}e_{ \acute{\gamma}}^{-}-
e_{\gamma}^{-}e_{ \acute{\gamma}}^{+}.
\eeqa
Thus the
general solution of the equation on the mass-shell is
\be\label{gensolution}
e_{\gamma\acute{\gamma}}=\xi_{\acute{\gamma}}k_{\gamma} +
\zeta_{\gamma} k_{\acute{\gamma}}+
c_{1}e^{+}_{\gamma\acute{\gamma}}+c_{2}e^{-}_{\gamma\acute{\gamma}}
+c_3 e^{A}_{\gamma\acute{\gamma}},
\ee
where $c_{1},c_{2}, c_{3}$ are arbitrary constants.
These are the propagating modes of {\it helicity-two and helicity-zero
$\lambda = \pm 2, 0$ charged gauge bosons}
\cite{Savvidy:2005fi,Savvidy:2005zm,Savvidy:2005ki}.
The propagating modes of higher rank gauge fields have been found in \cite{Savvidy:2009zz}.

\section{\it Invariant Forms for Fermions}

The fermions are defined as Rarita-Schwinger spinor-tensor fields
\cite{rarita,singh1,fronsdal1}
\be\label{raritaschwingerspinfields}
\psi^{\alpha}_{\lambda_1 ... \lambda_{s}}(x)
\ee
with mixed transformation properties of Dirac four-component wave
function  and are totally symmetric tensors of the rank $s$
over the indices $\lambda_1 ... \lambda_{s}$
(the index $\alpha$ denotes the Dirac index and will be
suppressed in the rest part of the article).
All fields of the $\{ \psi \}$ family
are isotopic multiplets  $\psi^{i}_{\lambda_1 ... \lambda_{s}}(x)$
belonging to the same representation $\sigma^{a}_{ij}$ of the compact Lie group G
(the index $i$ denotes the isotopic index).
One can think of these spinor-tensor fields as appearing in the
expansion of the extended fermion field $\Psi^{i}(x,e)$ over the unit
tangent vector
$e_{\lambda}$ \cite{Savvidy:2005fi,Savvidy:2005ki}:
\be\label{fermionfield}
\Psi^{i}(x,e) = \sum^{\infty}_{s=0}~
\psi^{i}_{\lambda_1 ... \lambda_{s}}(x) ~e_{\lambda_1}...e_{\lambda_s}.
\ee
Our intention is to introduce gauge invariant interaction of
fermion fields with non-Abelian tensor gauge fields.
The transformation of the fermions under the extended isotopic group we
shall define by the formula \cite{Savvidy:2005zm}
\beqa
  \Psi^{'}(x,e) &=& \CU(\xi) \Psi(x,e),
\eeqa
where
$
\CU(\xi) = \exp (i g \xi(x,e)) ,~~\xi(x,e)=   \sum^{\infty}_{s=0}~
\xi^{a}_{\lambda_1 ... \lambda_{s}}(x) ~\sigma^a   e_{\lambda_1}...e_{\lambda_s}
$
and $\sigma^{a}$ are the matrices of the
representation $\sigma$ of the compact Lie
group G, according to which all $\psi's$ are transforming. In components the transformation of
fermion fields under the extended isotopic group therefore will be \cite{Savvidy:2005zm}
\beqa\label{mattertransformation}
\delta_{\xi}  \psi  &=& i g \sigma^{a}\xi^{a}  \psi ,\nonumber\\
\delta_{\xi}  \psi_{\lambda} &=& i g \sigma^{a}( \xi^{a} ~\psi_{\lambda}   +
\xi^{a}_{\lambda}~ \psi) ,\nonumber\\
\delta_{\xi}  \psi_{\lambda\rho} &=& i g \sigma^{a}( \xi^{a} ~\psi_{\lambda\rho}   +
\xi^{a}_{\lambda}~ \psi_{\rho} + \xi^{a}_{\rho}~
\psi_{\lambda} + \xi_{\lambda\rho} ~ \psi),\\
........&.&.......................,\nn
\eeqa
The covariant derivative of the fermion field is defined as usually:
\be
\nabla_\mu \Psi = i \partial_\mu \Psi +g \CA_{\mu}(x,e) \Psi,
\ee
and  transforms  homogeneously:
$
\nabla_\mu \Psi \rightarrow \CU ~ \nabla_\mu \Psi ,
$
where we are using the matrix notation for the gauge fields
$\CA_{\mu} = \sigma^{a} \CA^{a}_{\mu}$. Therefore the gauge invariant
Lagrangian has the following form:
\be\label{generalfermionlagrangian}
\CL^F = \bar{\Psi} \gamma_{\mu} [i \partial_\mu \Psi +g \CA_{\mu}] \Psi.
\ee
Expanding this Lagrangian over the vector variable $e_{\lambda}$ one can get a
series of gauge invariant forms for half-integer fermion fields:
\be\label{firstlagrangian}
\CL^F = \sum^{\infty}_{s=0}   \CL_{s +1/2},
\ee
where $f_s$ are coupling constants.
The lower-spin invariant Lagrangian for the spin-1/2 field is:
\be\label{dirac}
{{\cal L}}_{1/2} =   \bar{\psi}^i  \gamma_{\mu} (\delta_{ij} i\partial_{\mu} ~+~
g \sigma^{a}_{ij} A^{a}_{\mu} )\psi^j = \bar{\psi}  ( i \not\!\partial +
g \not\!\!A )\psi
\ee
and for the spin-vector field $\psi_{\mu}$ together with the additional
rank-2 spin-tensor $\psi_{\mu\nu}$ the invariant Lagrangian has the form \cite{Savvidy:2005zm}:
\beqa\label{firstfermionlagrangianthreehalf}
{{\cal L}}_{3/2} &=&
\bar{\psi}_{\lambda} \gamma_{\mu} ( i\partial_{\mu} + g A_{\mu} )\psi_{\lambda} +
{1\over 2}\bar{\psi} \gamma_{\mu} (i\partial_{\mu} + g A_{\mu} )\psi_{\lambda\lambda}+
{1\over 2}\bar{\psi}_{\lambda\lambda} \gamma_{\mu}
(i\partial_{\mu} + g A_{\mu} )\psi\nonumber\\
&+& g \bar{\psi}_{\lambda} \gamma_{\mu}  A_{\mu\lambda} \psi
+g \bar{\psi}  \gamma_{\mu} A_{\mu\lambda} \psi_{\lambda}
+{1\over 2}g \bar{\psi} \gamma_{\mu}  A_{\mu\lambda\lambda} \psi ~.
\eeqa
As one can check it is invariant under simultaneous
gauge transformations of the fermions (\ref{mattertransformation}) and
tensor gauge fields (\ref{polygauge}):
$$
\delta {{\cal L}}_{3/2} =0.
$$
The Lagrangian (\ref{firstlagrangian}) is not the
most general Lagrangian which can be constructed in terms
of the above spinor-tensor fields (\ref{raritaschwingerspinfields}).
As we shall see, there exists a second invariant   ${{\cal L}}^{'}_{F}$
which can be constructed in terms of spinor-tensor
fields (\ref{raritaschwingerspinfields}),  and the total Lagrangian is a
linear sum: $  {{\cal L}}_F + f~ {{\cal L}}^{'}_F $.

Let us consider the gauge invariant tensor density of the form
\cite{Savvidy:2005fi,Savvidy:2005ki}
\be\label{generalfermiondensity}
{{\cal L}}_{\rho_1\rho_2} =    \bar{\Psi}(x,e)  \gamma_{\rho_1} [ i\partial_{\rho_2} ~+~
g \sigma^{a} \CA^{a}_{\rho_2}(x,e) ]\Psi(x,e) .
\ee
It is gauge invariant tensor density because its variation is equal to zero:
\beqa
\delta {{\cal L}}_{\rho_1\rho_2}(x,e) = i\bar{\Psi}(x,e)\xi(x,e)
\gamma_{\rho_1} [ i\partial_{\rho_2} ~+~
g  \CA_{\rho_2}(x,e) ]\Psi(x,e) +\nn\\
+ \bar{\Psi}(x,e)\gamma_{\rho_1}g (-{1\over g}) [\partial_{\rho_2}\xi(x,e)   -ig
[\CA_{\rho_2}(x,e), \xi(x,e) ]\Psi(x,e) +\nn\\
-i\bar{\Psi}(x,e)\gamma_{\rho_1} [ i\partial_{\rho_2} ~+~
g \sigma^{a} \CA^{a}_{\rho_2}(x,e) ]\xi(x,e)\Psi(x,e)=0,\nn
\eeqa
where $\CA_{\rho_2}(x,e)=\sigma^{a} \CA^{a}_{\rho_2}(x,e)$.
The Lagrangian density (\ref{generalfermiondensity}) generates the
series of {\it gauge invariant tensor densities
$(\CL^{'}_{\rho_1\rho_2})_{\lambda_1 ... \lambda_{s}}(x)$},
when we expand it in powers of the vector variable $e$:
\be\label{secondseriesdensities}
{{\cal L}}_{\rho_1\rho_2}(x,e) = \sum^{\infty}_{s=0}~ {1\over s!} ~
(\CL_{\rho_1\rho_2})_{\lambda_1 ... \lambda_{s}}(x) ~ e_{\lambda_1}...e_{\lambda_s} .
\ee
The gauge invariant tensor densities
$(\CL_{\rho_1\rho_2})_{\lambda_1 ... \lambda_{s}}(x)$ allow to construct
two series of gauge invariant forms:
$ {{\cal L}}_{s+1/2}$ and ${{\cal L}}^{'}_{s+1/2}~~$, s=1,2,.. by the
contraction of the corresponding tensor indices.
The lower gauge invariant tensor density has the form
\beqa\label{secondlowerspinlagrangian}
({{\cal L}}_{\rho_1\rho_2})_{\lambda_1\lambda_2}={1\over 2} \{~&+&
\bar{\psi}_{\lambda_1}   \gamma_{\rho_1} [ i\partial_{\rho_2} ~+~
g A_{\rho_2}  ]\psi_{\lambda_2}   +
\bar{\psi}_{\lambda_2}  \gamma_{\rho_1} [ i\partial_{\rho_2} ~+~
g A_{\rho_2}  ]\psi_{\lambda_1}  + \nonumber\\
&+&\bar{\psi}_{\lambda_1\lambda_2}    \gamma_{\rho_1} [ i\partial_{\rho_2} ~+~
g A_{\rho_2}  ]\psi  +
\bar{\psi}   \gamma_{\rho_1} [ i\partial_{\rho_2} ~+~
g A_{\rho_2}  ]\psi_{\lambda_1\lambda_2}  + \nonumber\\
&+&g \bar{\psi}_{\lambda_1}  \gamma_{\rho_1} A_{\rho_2\lambda_2}  \psi   +
g\bar{\psi}_{\lambda_2}    \gamma_{\rho_1} A_{\rho_2\lambda_1}  \psi   +\nonumber\\
&+&g \bar{\psi}  \gamma_{\rho_1} A_{\rho_2\lambda_2}  \psi_{\lambda_1}   +
g\bar{\psi}   \gamma_{\rho_1} A_{\rho_2\lambda_1} \psi_{\lambda_2}
+ g \bar{\psi}    \gamma_{\rho_1} A_{\rho_2\lambda_1\lambda_2} \psi \},
\eeqa
and we shall use it to generate Lorentz invariant densities. Performing contraction of
the indices of this tensor density with respect to $\eta_{\rho_1\rho_2}\eta_{\lambda_1\lambda_2}$
we shall reproduce our first gauge invariant Lagrangian
density ${\cal L}_{3/2}$ (\ref{firstfermionlagrangianthreehalf}) presented in the previous
section.
We shall get the second gauge invariant Lagrangian performing the
contraction with respect to the $\eta_{\rho_1\lambda_1}\eta_{\rho_2\lambda_2}$,
which is obviously different form the previous one:
\beqa\label{secondfermionlagrangianthreehalf}
{{\cal L}}^{'}_{3/2} &=&{1\over 2}\{
\bar{\psi}_{\mu} \gamma_{\mu} ( i\partial_{\lambda} + g A_{\lambda} )\psi_{\lambda} +
\bar{\psi}_{\lambda} ( i\partial_{\lambda} + g A_{\lambda} )\gamma_{\mu} \psi_{\mu} +\nn\\
&+& \bar{\psi}_{\mu\lambda} \gamma_{\mu} ( i\partial_{\lambda} + g A_{\lambda} )\psi +
\bar{\psi} ( i\partial_{\lambda} + g A_{\lambda} )\gamma_{\mu} \psi_{\mu\lambda} +\\
&+& g \bar{\psi}_{\mu} \gamma_{\lambda}  A_{\mu\lambda} \psi
+g \bar{\psi}  \gamma_{\mu} A_{\lambda\mu} \psi_{\lambda}
+g \bar{\psi}_{\mu} \gamma_{\mu}  A_{\lambda\lambda} \psi
+g \bar{\psi}  \gamma_{\mu} A_{\lambda\lambda} \psi_{\mu}
+g \bar{\psi} \gamma_{\mu}  A_{\lambda\mu\lambda} \psi ~\}.\nn
\eeqa
Independently, one can check
that the last expression is invariant under simultaneous
extended gauge transformations of fermions (\ref{mattertransformation}) and
tensor gauge fields (\ref{polygauge}), calculating its variation:
$$
\delta  \CL^{'}_{3/2} =0.
$$
As one can see  from  (\ref{firstfermionlagrangianthreehalf})
and (\ref{secondfermionlagrangianthreehalf}) the interaction of fermions with
tensor gauge bosons is going through the
cubic vertex which includes two fermions and a tensor gauge boson, very
similar to the vertices in QED and the Yang-Mills theory.

Thus the gauge invariant Lagrangian for fermions
contains two infinite series of invarinat forms
${{\cal L}}_{s +1/2}$,~${{\cal L}}^{'}_{s+1/2 }$
and is a linear sum of these forms
\be\label{fermiforms}
{{\cal L}}^{F} ={{\cal L}}_{1/2}~+ \sum^{\infty}_{s=1}~  ({{\cal L}}_{s +1/2}~+
 ~ f_{ s }~ {{\cal L}}^{'}_{s+1/2 }).
\ee
It is important to notice that the invariance with respect to
the extended gauge transformations does not fix the coupling constants
$f_{s} $. The coupling constants $f_{s} $
remain arbitrary because every term of the sum is separately gauge invariant
and the extended gauge symmetry alone does not define them.
{\it The basic principle which we shall pursue
in our construction will be to fix these coupling constants demanding
unitarity of the theory}\footnote{
For that one should study the spectrum of the theory and its dependence
on these coupling constants. For some particular values of coupling constants
the linear sum of these forms  may exhibit symmetries with respect to a
bigger gauge group securing the absence of ghost states.}.

\section{\it Propagating Modes of Tensor Fermions}
As we have seen above the  Lagrangian for lower rank fields is a
linear sum
\beqa\label{totalfermionlagrangianthreehalf}
\CL={{\cal L}}_{1/2} + {{\cal L}}_{3/2} +f_1 {{\cal L}}^{'}_{3/2}+...  \nn
\eeqa
and our aim  is to find out the coefficient $f_1$  for
which the equation for the spin-vector field $\psi_{\lambda}$
correctly describes  the propagation of spin 3/2.
In the limit $g \rightarrow 0$ the kinetic part of the Lagrangian will take the form
\beqa\label{freefermionlagrangianthreehalf}
{{\cal L}}_{3/2} +f_1 {{\cal L}}^{'}_{3/2} \vert_{kinetic} &=&
\bar{\psi}_{\lambda} \gamma_{\mu}  i\partial_{\mu} \psi_{\lambda} +
{1\over 2}\bar{\psi} \gamma_{\mu} i\partial_{\mu} \psi_{\lambda\lambda}+
{1\over 2}\bar{\psi}_{\lambda\lambda} \gamma_{\mu}
i\partial_{\mu}\psi\nonumber\\
 &+&{f_1 \over 2}\{
\bar{\psi}_{\mu} \gamma_{\mu}  i\partial_{\lambda} \psi_{\lambda} +
\bar{\psi}_{\lambda}  i\partial_{\lambda} \gamma_{\mu} \psi_{\mu} +
\bar{\psi}_{\mu\lambda} \gamma_{\mu} i\partial_{\lambda}\psi +
\bar{\psi}  i\partial_{\lambda} \gamma_{\mu} \psi_{\mu\lambda} ~\} \nn
\eeqa
and we have the following free equation of motion for the spin-vector field $\psi_{\lambda}$:
\beqa\label{freefermionequationsshort}
\!\not\!p \psi_{\mu} +
f_1{1\over 2}  (\gamma_{\mu} p_{\lambda}
+ \gamma_{\lambda}p_{\mu})\psi_{\lambda}&=&0,\nn
\eeqa
where $\!\not\!p = \gamma_{\mu} p_{\mu}= \gamma_{\mu} i\partial_{\mu} $.
Let us consider the gauge transformation of the
spin-vector field $\psi_{\lambda}$ of the Rarita-Schwinger form:
\beqa\label{enhancedfermiontransformation}
\delta \psi_{\lambda } &=& \partial_{\lambda } \varepsilon~.
\eeqa
The variation of the  equation shows that
for $f_1=-2$ the equation
is invariant with respect to the RS transformation, if the spinor  parameter
$\varepsilon $ fulfils the condition
\be
\partial^2 \varepsilon  =0.
\ee
The equation for the spin-vector field $\psi_{\lambda}$ therefore is
\beqa\label{freefermion3/2}
H_{\mu\lambda} u_{\lambda}=(\!\not\!p \eta_{\mu\lambda }- \gamma_{\mu} p_{\lambda}
- \gamma_{\lambda}p_{\mu}) u_{\lambda}=
0,
\eeqa
where  the matrix operator is
$
H_{\mu\lambda}= \eta _{\mu\lambda} \!\not\!p - \gamma_{\mu} p_{\lambda}
- \gamma_{\lambda}p_{\mu}
$
and  $\psi_{\lambda} = u_{\lambda} \exp {(ipx)}$.

Let us first consider the Dirac equation for the spinor $\psi =u \exp(ipx)$
and the Lagrangian $\CL_{1/2}$ (\ref{dirac}):
$$
\!\not\!p u =0~.
$$
It describes particles of helicities $\lambda=\pm 1/2$, so that the wave functions are
$$
\omega=+p,~~
u_+=  \left(\begin{array}{cccc}
0 \\
0 \\
1 \\
0 \\
\end{array} \right),~u_-= \left(\begin{array}{cccc}
0 \\
1 \\
0 \\
0 \\
\end{array} \right),~~~\omega=-p,~~
v_+=  \left(\begin{array}{cccc}
1 \\
0 \\
0 \\
0 \\
\end{array} \right),~v_-= \left(\begin{array}{cccc}
0 \\
0 \\
0 \\
1 \\
\end{array} \right),
$$
where we have chosen the momentum vector
in the third direction  $p^{\mu} = (\omega, 0,0,p)$.

Now let us consider the matrix operator
$H_{\mu\lambda}$ for spin-vector $\psi_{\lambda}$.
If $\omega^2 - p^2 \neq 0$, the rank of the 16-dimensional matrix
$H_{\mu\lambda}$ is $rank~H\vert_{\omega^2 - p^2 \ne 0}=16$
and we have only trivial solution $u_{\lambda} = 0$.
If $\omega = + p $,
then the rank of the matrix drops, $rank~H\vert_{\omega= +p}=10$,
and the number of independent solutions is 16-10=6.
These six solutions of the equation (\ref{freefermion3/2}) are
\beqa
u^{\alpha ~(gauge)}_{\lambda} =p_{\lambda} ~\otimes ~\varepsilon^{\alpha},~~~~~
  u^{\alpha}_{\lambda} = e_{\mu}^{\pm } ~\otimes ~u^{\alpha}_{\pm}
\eeqa
where  $\varepsilon^{\alpha}$ is a spinor gauge parameter.
The first four solutions are pure gauge fields ($\sim p_{\lambda}$),
while the remaining two are the physical modes of helicities $\lambda=\pm 3/2$.
If $\omega = - p $ we have again  gauge modes and two
physical modes of helicities $\lambda=\pm 3/2$  describing antiparticles.
The general solution on the mass-shell
will be a linear combination of all these solutions.
We conclude that equation (\ref{freefermion3/2}) correctly describes the
free propagation of physical modes of the massless particle of spin 3/2.

In summary we have the following massless spectrum for the lower rank boson and fermion fields:
\beqa
A_{\mu}:~~~~\lambda=\pm 1 ~~~~~~~~~~~~\psi:~~~~\lambda=\pm 1/2, \nn\\
A_{\mu\nu}:~~~~\lambda=\pm 2,0 ~~~~~~~~\psi_{\mu}:~~~~\lambda=\pm 3/2,\\
.....................................................................\nn
\eeqa
The propagating modes of higher rank gauge fields have been analyzed in \cite{Savvidy:2009zz}.

\section{\it Topological Mass Generation}

Several mechanisms are currently known for generating massive vector particles that
are compatible with the gauge invariance. One of them
is the spontaneous symmetry breaking mechanism, which
generates masses and requires the existence of
the fundamental scalar particle - the Higgs boson. The scalar field
provides the longitudinal polarization of the massive vector boson
and ensures unitarity of its scattering
amplitudes \cite{Cornwall:1973tb,Llewellyn Smith:1973ey}\footnote{
Extended discussion and references can be found in
\cite{Schwinger:1962tn,Schwinger:1962tp,Goto:1967,Veltman:1968ki,Slavnov:1970tk,
vanDam:1970vg,Slavnov:1972tk,Veltman:2000xp,Sikivie:1980fm,
Farhi:1980xs,Dimopoulos:1980fj,Slavnov:2006rf}.}.

The argument in favor of a pure gauge field theory mechanism
was a dynamical mechanism of mass generation proposed  by Schwinger
\cite{Schwinger:1962tn}, who was arguing that the gauge
invariance of a vector field does not necessarily lead to the massless spectrum
of its excitations and suggested its realization in (1+1)-dimensional gauge
theory\cite{Schwinger:1962tp}.

Compatibility of gauge invariance and mass term in
(2+1)-dimensional gauge field theory was demonstrated by
Deser, Jackiw and Templeton \cite{Deser:1982vy, Deser:1981wh} and
Schonfeld \cite{Schonfeld:1980kb}, who added to
the YM Lagrangian a gauge invariant Chern-Simons density.

Here we suggest a similar mechanism that generates masses of the YM boson
and tensor gauge bosons in (3+1)-dimensional space-time at the classical level
\cite{Savvidy:2010bk}. As we shall
see, in  non-Abelian tensor gauge theory \cite{Savvidy:2005fi,Savvidy:2005zm,Savvidy:2005ki}
there exists a gauge invariant,
metric-independent density $\Gamma $ in five-dimensional
space-time which is the derivative of the vector current $\Sigma_{\mu}$.
This invariant in five dimensions has many properties
of the Chern-Pontryagin density $\CP= \partial_{\mu} C_{\mu}$ in
four-dimensional YM theory, which  is
a derivative of the Chern-Simons topological vector current $C_{\mu}$.
The fifth component of the vector
current $\Sigma_{4}  \equiv \Sigma $  is a
gauge invariant density  which is defined in four-dimensional
space-time. Its dimensionality is $[mass]^3$. Adding this term  to the Lagrangian
of non-Abelian tensor gauge fields leaves intact its gauge
invariance and generates massive vector particles of the  mass
\be
M^2 =  {4 \over 3 } ~m^2 .
\ee
A massive spin-1  particle appears here as a vector field of helicities
$\lambda=\pm 1$ which acquires a third longitudinal polarization
absorbing antisymmetric field which carries a zero helicity $\lambda=0$ (\ref{helicities}).

Let us consider this new invariant in five-dimensional space-time $(4+1)$,
which can be constructed by means of the
totaly antisymmetric Levi-Civita epsilon tensor $\varepsilon_{\mu\nu\lambda\rho\sigma}$
($\mu,\nu,...=0,1,2,3,4$) in combination with the generalized field strength tensors
(\ref{fieldstrengthparticular}) \cite{Savvidy:2010bk}
\beqa\label{freeactionthreeprimesum}
\Gamma
= \varepsilon_{\mu\nu\lambda\rho\sigma} TrG_{\mu\nu}G_{\lambda\rho, \sigma}
=2~\varepsilon_{\mu\nu\lambda\rho\sigma}G^{a}_{\mu\nu}G^{a}_{\lambda\rho, \sigma}.
\eeqa
We shall demonstrate that this invariant in five dimensions has many properties
of the Chern-Pontryagin density
\be\label{chernpontragyn}
\CP= {1\over 4}\varepsilon_{\mu\nu\lambda\rho }Tr G_{\mu\nu}G_{\lambda\rho }=\partial_{\mu} C_{\mu}
\ee
in Yang-Mill theory in four dimensions, where
\be\label{chernsimons}
C_{\mu}=\varepsilon_{\mu\nu\lambda\rho }Tr (A_{\nu}\partial_{\lambda}A_{\rho }
-i{2\over 3}g  A_{\nu}A_{\lambda}A_{\rho })
\ee
is the Chern-Simons topological current.
Indeed, $\Gamma$ is obviously diffeomorphism-invariant and does not involve a
space-time metric.
It is gauge invariant because under the gauge transformation $\delta_{\xi} $
(\ref{polygauge}) it vanishes:
\beqa
\delta_{\xi} \Gamma
&=&-i g  \varepsilon_{\mu\nu\lambda\rho\sigma} Tr ( [G_{\mu\nu} ~\xi] G_{\lambda\rho, \sigma}
 +G_{\mu\nu} (~[~G_{\lambda\rho,\sigma}~ \xi ]
+  [G_{\lambda\rho} ~\xi_{\sigma}]~))=0.\nn
\eeqa
The variation of its integral over the gauge fields
$A^{a}_{\mu}$ and $A^{a}_{\mu\lambda}$ gives:
\beqa\label{topologicalinvariant}
\delta_{A} \int_{M_5} d^5x ~\Gamma &=&
-  2\varepsilon_{\mu\nu\lambda\rho\sigma} \int d^5x Tr
((\nabla_{\mu} G_{\lambda\rho, \sigma}
- i g [A_{\mu\sigma} ~G_{\lambda\rho}])~\delta A_{\nu}
+(\nabla_{\lambda} G_{\mu\nu}) ~\delta A_{\rho\sigma})\nn\\
&+& 2 \varepsilon_{\mu\nu\lambda\rho\sigma} \int d^5x Tr
(\nabla_{\mu} (G_{\lambda\rho, \sigma} ~\delta A_{\nu})
+\nabla_{\lambda} (G_{\mu\nu} ~\delta A_{\rho\sigma})) .\nn
\eeqa
Recalling the Bianchi identity in the YM theory
and the generalized Bianchi identities for higher-rank  field strength tensor $G_{\nu\lambda,\rho}$
presented in the Appendix, one can see that  $\Gamma$  gets contribution
only from the boundary terms and vanishes when the fields vary in the
bulk of the manifold\footnote{The trace of the commutators vanishes:
$Tr([A_{\mu} ;G_{\lambda\rho, \sigma} \delta A_{\nu}]
+[A_{\lambda};  G_{\mu\nu} \delta A_{\rho\sigma}])=0$.}:
\beqa
\delta_{A} \int_{M_5} d^5x ~\Gamma =  2 \varepsilon_{\mu\nu\lambda\rho\sigma}
\int_{M_5}  d^5x ~\partial_{\mu} ~Tr
(G_{\lambda\rho, \sigma}~\delta A_{\nu}
+G_{\nu\lambda} ~\delta A_{\rho\sigma})=\nn
\\
= 2 \varepsilon_{\mu\nu\lambda\rho\sigma}
\int_{\partial M_5}~Tr
(G_{\lambda\rho, \sigma}~\delta A_{\nu}
+G_{\nu\lambda} ~\delta A_{\rho\sigma}) d\sigma_{\mu} =0.\nn
\eeqa
Therefore  $\Gamma$  is insensitive
to the {\it local} variation of the fields.
It becomes obvious that $\Gamma$ is a total derivative of some
vector current $\Sigma_{\mu}$. Indeed, simple algebraic computation gives
$
\Gamma
=\varepsilon_{\mu\nu\lambda\rho\sigma} TrG_{\mu\nu}G_{\lambda\rho, \sigma}=
\partial_{\mu} \Sigma_{\mu},
$
where
\beqa\label{topologicalcurrent}
\Sigma_{\mu}
&=& 2 \varepsilon_{\mu\nu\lambda\rho\sigma} Tr (A_{\nu} ~\partial_{\lambda} A_{\varrho\sigma }
-  \partial_{\lambda} A_{\nu} ~ A_{\rho\sigma }
- 2 i g A_{\nu} A_{\lambda} A_{\rho\sigma }).
\eeqa
After some rearrangement and taking into account the definition of the field strength
tensors (\ref{fieldstrengthparticular}) we can get the following
form of the vector current \cite{Savvidy:2010bk}:
\beqa\label{topologicalcurrent1}
\Sigma_{\mu}
&=&  \varepsilon_{\mu\nu\lambda\rho\sigma} Tr G_{\nu\lambda }A_{\rho  \sigma}.
\eeqa
It is instructive  to compare the expressions (\ref{freeactionthreeprimesum}), (\ref{chernpontragyn})
and (\ref{chernsimons}), (\ref{topologicalcurrent1}). Both entities $\CP$
and $\Gamma$ are
metric-independent, are insensitive to the local variation of the fields and are derivatives of the
corresponding vector currents $C_{\mu}$ and $\Sigma_{\mu}$. The difference between them is that the former is
defined in four dimensions, while the latter in five. This difference in one unit of the space-time
dimension originates from the fact that we have at our disposal high-rank tensor gauge fields
to build new invariants.
The same is true for the Chern-Simons topological current $C_{\mu}$ and for
the current $\Sigma_{\mu}$, where
the latter is defined in five dimensions.
It is also remarkable that the current  $\Sigma_{\mu}$ is linear in the YM field strength tensor
and in the rank-2 gauge field, picking up only its antisymmetric part.

While the invariant $\Gamma$ and the vector current $\Sigma_{\mu}$ are defined on a five-dimensional
manifold, we may restrict the latter to one lower, four-dimensional manifold. The
restriction proceeds as follows.
Let us consider the fifth component of the vector current $\Sigma_{\mu}$:
\beqa\label{topologicalcharge}
\Sigma \equiv \Sigma_{4}
&=&  \varepsilon_{4\nu\lambda\rho\sigma} Tr G_{\nu\lambda }A_{\rho  \sigma}.
\eeqa
Considering the fifth component of the vector
current $\Sigma \equiv\Sigma_4$ one can see that the remaining indices will not repeat
the external index and the sum is restricted to the sum over indices of four-dimensional
space-time.
Therefore we can reduce this functional to four dimensions.
This is the case when the gauge fields are independent
on the fifth coordinate $x_4$. Thus the density $\Sigma$ is well
defined in four-dimensional space-time and, as we shall see,
it is also gauge invariant up to the total divergence term.  Therefore we shall
consider its integral over four-dimensional space-time\footnote{Below we
are using the same
Greek letters to numerate now the four-dimensional coordinates. There should be
no confusion because the
dimension  can always be recovered from  the dimension of the epsilon tensor.}:
\beqa\label{topologicalSigmaCS}
\int_{M_4} d^4 x~ \Sigma
&=&  \varepsilon_{\nu\lambda\rho\sigma} \int_{M_4} d^4 x~ Tr~ G_{\nu\lambda } A_{\rho  \sigma} .
\eeqa
This entity is an analog of the Chern-Simon secondary characteristic
\be\label{chernsimonscharcterictic}
CS =\varepsilon_{ijk } \int_{M_3}  d^3x ~Tr~ (A_{i}\partial_{j}A_{k }
-i g {2\over 3} A_{i}A_{j}A_{k }),
\ee
but, importantly, instead of being defined in three dimensions
it is now  defined in four dimensions. Thus the non-Abelian tensor gauge fields
allow to build a natural generalization of the Chern-Simons characteristic in four-dimensional
space-time.

As we claimed this functional is gauge invariant up to the total divergence term.
Indeed, its gauge variation under $\delta_\xi$ (\ref{polygauge}) is
\beqa\label{variation}
\delta_{\xi} \int_{M_4} d^4 x~ \Sigma
&=& \varepsilon_{\nu\lambda\rho\sigma} \int_{M_4} Tr (-i g [G_{\nu\lambda }~\xi] A_{\rho  \sigma}
+ G_{\nu\lambda } (\nabla_{\rho} \xi_{\sigma}-i g  [A_{\rho  \sigma}~\xi] )) d^4 x =\nn\\
&=&\varepsilon_{\nu\lambda\rho\sigma} \int_{M_4} \partial_{\rho} ~Tr (
G_{\nu\lambda }  \xi_{\sigma} ) d^4 x=\varepsilon_{\nu\lambda\rho\sigma}
\int_{\partial M_4} Tr (
G_{\nu\lambda }  \xi_{\sigma} ) d\sigma_{\rho} =0 .
\eeqa
Here the first and the third  terms cancel each other and the second one, after integration by
part and recalling the Bianchi identity  (\ref{newbianchi}),  leaves only the
boundary term, which vanishes  when
the gauge parameter $\xi_{\sigma}$ tends to zero at infinity.

It is interesting to know whether the invariant $\Sigma$ is associated with some
new topological characteristic of the gauge fields.
If the YM field strength $G_{\nu\lambda }$ vanishes, then the vector potential is equal to the
pure gauge connection $A_{\mu}=U^{-}\partial_{\mu} U$. Inspecting the expression for the
invariant $\Sigma$ one can get convinced that it vanishes on such fields because there is
a field strength tensor $G_{\nu\lambda }$ in the integrant. Therefore
it does not differentiate topological properties of the gauge function $U$,
like its winding number. Both "small" and "large" gauge transformations have zero
contribution to this invariant. It may distinguish fields which are
falling less faster at infinity and have nonzero field strength tensor $G_{\nu\lambda }$
and the tensor gauge field $A_{\rho  \sigma} $.

In four dimensions the gauge fields have dimension of $[mass]^1$,
therefore if we intend to add this new density to the Lagrangian we should introduce the mass parameter $m$:
\beqa\label{topologicalmass}
m~\Sigma
&=&  m~\varepsilon_{\nu\lambda\rho\sigma}   Tr ~G_{\nu\lambda } A_{\rho  \sigma} ,
\eeqa
where parameter $m$ has units $[mass]^1$. Adding this term  to the Lagrangian
of non-Abelian tensor gauge fields keeps intact its gauge
invariance and our aim is to analyze the particle spectrum of this gauge field theory.
The natural appearance of the mass parameters
hints at the fact that the theory turns out to be a massive theory. We shall see
that the YM vector boson becomes massive, suggesting an alternative
mechanism for mass generation
in gauge field theories in four-dimensional space-time.

We have to notice that the Abelian version of the invariant $\Sigma$ was investigated earlier
in \cite{Cremmer:1973mg,Hagen:1978jk,Kalb:1974yc,Nambu:1975ba,Ogievetsky:1967ij,Aurilia:1981xg,
Freedman:1980us,Slavnov:1988sj,Allen:1990gb}.
Indeed, if one considers instead of a non-Abelian group the Abelian group one can see
that the invariant $\Sigma$ reduces to the
$\varepsilon_{\nu\lambda\rho\sigma}   F_{\nu\lambda } B_{\rho  \sigma}$ and when
added to the Maxwell Lagrangian provides a  mass to the vector field
\cite{Cremmer:1973mg,Kalb:1974yc,Ogievetsky:1967ij,Nambu:1975ba,Aurilia:1981xg,
Freedman:1980us}.
Attempts at producing a non-Abelian  invariant
in a similar way  have come up with difficulties  because they
involve non-Abelian generalization of
gauge transformations of antisymmetric fields
\cite{Freedman:1980us,Slavnov:1988sj,Henneaux:1997mf,Lahiri:1996dm,Botta Cantcheff:2003cv}.
Let us compare the
formulas (2.16) and (2.17) suggested in \cite{Freedman:1980us,Slavnov:1988sj}
for the transformation of antisymmetric field with the gauge transformation
$\delta_{\xi}$  (\ref{polygauge}).
For lower-rank fields the latter can be written in the following way:
\beqa
\delta_{\xi}  A_{\mu} &=& \partial_{\mu}\xi -i g[A_{\mu},\xi],~~~~~~
\delta_{\xi}  A_{\mu\nu} =  -i g [A_{\mu\nu},\xi],
\nonumber\\
\delta_{\zeta}  A_{\mu} &=& 0,~~~~~~~~~~~~~~~~~~~~~~~
\delta_{\zeta}  A_{\mu\nu} = \partial_{\mu}\zeta_{\nu} -i g[A_{\mu},\zeta_{\nu}] .\nonumber
\eeqa
The  antisymmetric part of this transformation amazingly coincides with the one
suggested  in \cite{Freedman:1980us} if one takes the auxiliary field $A^i_{\mu}$ of
\cite{Freedman:1980us} equal to zero.
The crucial point is that the gauge transformations of non-Abelian tensor gauge fields
\cite{Savvidy:2005fi,Savvidy:2005zm,Savvidy:2005ki}
defined in (\ref{polygauge}) cannot be limited to a YM vector and antisymmetric
field $B^{a}_{\mu\nu}$. Instead, antisymmetric field is
augmented by a symmetric rank-2 gauge field, so that together they form a
gauge field $A^{a}_{\mu\nu}$ which transforms as it is given above and is a fully propagating
field.
It is also important that one should include all high-rank gauge fields in order to be able
to close the group of gauge transformations and to construct
invariant Lagrangian.

Let us now see
how the spectrum is changing when we add new invariant  $\Sigma$
(\ref{topologicalmass}) to the Lagrangian.
With the new mass term  the Lagrangian takes the form
\beqa
\CL  = \CL_{YM} +    \CL_{2}+ \CL^{'}_{2}  + {m \over 4} ~\Sigma~ .
\eeqa
The free equations (g=0) of motion for the YM and rank-2 gauge fields
are:
\beqa
 \partial^2 A^{a}_{\nu  } -
\partial_{\nu} \partial_{\mu} A^{a}_{\mu }  +
m ~\varepsilon_{\nu \mu\lambda\rho} \partial_{\mu} A^{a}_{\lambda\rho}= 0,~~~~~~~~~~~~~~~~~~~~~\nn\\
 \partial^{2}(A^{a}_{\nu\lambda} -{1\over 2}A^{a}_{\lambda\nu})
-\partial_{\nu} \partial_{\mu}  (A^{a}_{\mu\lambda}-
{1\over 2}A^{a}_{\lambda\mu} )-
\partial_{\lambda} \partial_{\mu}  (A^{a}_{\nu\mu} - {1\over 2}A^{a}_{\mu\nu} )
+~~~~\nn\\
 +\partial_{\nu} \partial_{\lambda} ( A^{a}_{\mu\mu}-{1\over 2}A^{a}_{\mu\mu})
+{1\over 2}\eta_{\nu\lambda} ( \partial_{\mu} \partial_{\rho}A^{a}_{\mu\rho}
-  \partial^{2}A^{a}_{\mu\mu}) +
 m ~\varepsilon_{\nu\lambda\mu \rho} \partial_{\mu} A^{a}_{\rho} = 0.
\eeqa
This is a coupled system of equations which involve the vector YM field
and the antisymmetric part of the rank-2 gauge field.
Only the antisymmetric part $B_{\nu\lambda}$
of the rank-2 gauge field $A_{\nu\lambda}$ interacts through the
mass term, the symmetric part $A^S_{\nu\lambda}$ completely decouples from
both equations\footnote{The symmetric field
can acquire a mass when we include the next invariant mass term
$m_3~\Sigma_3$ \cite{Savvidy:2010bk}.}, therefore
we arrive at the following system of equations:
\beqa\label{systeminmomentum}
 (-k^2 \eta_{\nu \mu }+
k_{\nu} k_{\mu} ) e_{\mu }  +
 i m ~\varepsilon_{\nu \mu\lambda\rho} k_{\mu} b_{\lambda\rho}= 0,~~~~~~~~~~\nn\\
(- k^{2} \eta_{\nu \mu } \eta_{\lambda\rho}
+k_{\nu} k_{\mu}\eta_{\lambda\rho}
-\eta_{\nu \mu } k_{\lambda} k_{\mu} ) b_{\mu\rho}  +
i {2 m \over 3}  ~\varepsilon_{\nu\lambda\mu \rho} k_{\mu} e_{\rho} = 0.
\eeqa
When $k^2 \neq M^2$ the system (\ref{systeminmomentum}) is off mass-shell and we have
four  pure gauge field solutions:
\beqa\label{puregauge}
e_{\mu}= k_{\mu},~~~~b_{\nu\lambda} = 0;~~~~~~~~~~~~~~~~
e_{\mu}=0,~~~~b_{\nu\lambda} =  k_{\nu}\xi_{\lambda}-k_{\lambda}\xi_{\nu}.
\eeqa
When $k^2 \neq M^2$ the system (\ref{systeminmomentum}) has seven solutions. These are four
pure gauge solutions (\ref{puregauge}) and additional three solutions representing
propagating modes:
\beqa\label{topphysicalmodes}
e^{(1)}_{\mu}= (0,1,0,0),~~
b_{\gamma\acute{\gamma}}^{(1)}={1\over i}{M \over \sqrt{\vec{k}^2 + M^2}}
\left(\begin{array}{cccc}
0&0&0&0 \\
0&0&0&0 \\
0&0&0&1 \\
0&0&-1&0 \\
\end{array} \right),
\nn\\
e^{(2)}_{\mu}= (0,0,1,0),~~
b_{\gamma\acute{\gamma}}^{(2)}=-{1\over i}{M \over \sqrt{\vec{k}^2 + M^2}}
\left(\begin{array}{cccc}
0&0&0&0 \\
0&0&0&1 \\
0&0&0&0\\
0&-1&0&0 \\
\end{array} \right),\nn\\
e^{(3)}_{\mu}= (0,0,0,{M \over \sqrt{\vec{k}^2 + M^2}}),~~
b_{\gamma\acute{\gamma}}^{(3)}={1\over i}
\left(\begin{array}{cccc}
0&0&0&0 \\
0&0&1&0\\
0&-1&0&0 \\
0&0&0&0 \\
\end{array} \right).
\eeqa
These propagating modes cannot be factorized into separately vector
or separately tensor solutions as it happens  for the pure
gauge solutions (\ref{puregauge}). It is a genuine superposition of
vector and tensor fields.
Let us consider the limit $M \rightarrow 0$. The above solutions will
factorize  into two massless vector modes $ e^{(1)}_{\mu},~
e^{(2)}_{\mu},~$ of helicities $\lambda = \pm 1$
and helicity $\lambda = 0$ mode $b_{\gamma\acute{\gamma}}^{(3)}
$ of antisymmetric tensor. But when $M \neq 0$, in the rest
frame $\vec{k}^2 =0$, these solutions represent three polarizations of the spin-1
boson.

The above analysis suggests the following physical interpretation.
A massive spin-1  particle appears here as a vector field of helicities
$\lambda=\pm 1$ which acquires an extra polarization state absorbing
antisymmetric field of helicity $\lambda=0$, or as antisymmetric field of helicity  $\lambda=0$
which absorbs helicities $\lambda=\pm 1$ of the vector field.
It is sort of "dual" description of massive spin-1 particle.
In order to fully justify this phenomenon of superposition of polarizations
one should develop quantum-mechanical description of tensor fields.
There is a need for deeper understanding
of the corresponding  path integral which is over infinitely many fields.

In conclusion,  I wish to thank the organizers of the conference
for the invitation and for arranging an interesting
and stimulating meeting. I also would like to thank Irina Aref'eva,
Ludwig Faddeev and  Andrey Slavnov
for interesting discussions during the conference.

\section{\it Appendix}
The field strength tensors fulfil the Bianchi
identities. In YM theory it is
\be
[\nabla_{\mu},G_{\nu\lambda}]+[\nabla_{\nu},G_{\lambda\mu}]+
[\nabla_{\lambda},G_{\mu\nu}]=0,
\ee
for the higher rank field strength tensors $G_{\nu\lambda,\rho}$ and
$G_{\nu\lambda,\rho\sigma}$ they are:
\be\label{newbianchi}
[\nabla_{\mu},G_{\nu\lambda,\rho}]-ig[A_{\mu\rho},G_{\nu\lambda}]+
[\nabla_{\nu},G_{\lambda\mu,\rho}]-ig[A_{\nu\rho},G_{\lambda\mu}]+
[\nabla_{\lambda},G_{\mu\nu,\rho}]-ig[A_{\lambda\rho},G_{\mu\nu}]=0,
\ee
\be\label{newbianchi3}
[\nabla_{\mu},G_{\nu\lambda,\rho\sigma}]-ig[A_{\mu\rho},G_{\nu\lambda,\sigma}]
-ig[A_{\mu\sigma},G_{\nu\lambda,\rho}]-ig[A_{\mu\rho\sigma},G_{\nu\lambda}]
+ cyc.perm. (\mu\nu\lambda)=0
\ee
and so on.

\vfill
\end{document}